\magnification=\magstep1
\baselineskip=18pt
\hfuzz=6pt

$ $

\vskip 1cm

\centerline{\bf Quantum computation with abelian anyons}

\bigskip

\centerline{Seth Lloyd}

\smallskip

\centerline{Department of Mechanical Engineering}

\centerline{MIT 3-160, Cambridge, Mass. 02139}

\centerline{slloyd@mit.edu}

\bigskip
\noindent{\it Abstract:} A universal quantum computer
can be constructed using abelian anyons.  
Two qubit quantum logic gates such as controlled-NOT operations
are performed using topological effects.  Single-anyon
operations such as hopping from site to site on a lattice
suffice to perform all quantum logic operations.  
Quantum computation using abelian anyons shares some but not
all of the robustness of quantum computation using non-abelian 
anyons.

\bigskip
A wide variety of methods can be used to construct quantum computers
in principle (1-18).  Essentially any interaction between two quantum degrees
of freedom suffices to construct universal quantum logic gates (10-11).
In particular, Kitaev (12) has shown that quantum computation can 
be effected using non-abelian anyons.  The resulting quantum computation
is intrinsically fault-tolerant. This paper shows that universal quantum
computation can be effected using abelian anyons: two-qubit
quantum logic gates, such as the controlled-NOT gate, are enacted
topologically via the phase factor of $e^{i\phi}$ that occurs when one
anyon is moved around another.  Quantum computation using abelian
anyons shares some but not all of the robustness of quantum computation
using non-abelian anyons.  Possible realizations 
in terms of solid-state systems and interferometers
are discussed, and issues of noise and decoherence are investigated. 

A quantum logic gate is an operation that transforms quantum-information
bearing degrees of freedom.
A set of quantum logic gates is universal if arbitrary quantum
computations can be built up by repeatedly applying gates from the
set to different qubits.  To prove that abelian anyons
are capable of universal quantum computation, we  will show
that single anyon quantum logic gates such as those that move an anyon from 
one site to another form a universal set: universal quantum computation can be 
effected simply by moving fermions around a lattice or network.  

Consider the case of anyons moving on
a two-dimensional lattice.  Each site $j$ of the lattice corresponds
to a local mode that can be either occupied by an anyon, $|+\rangle_j$,
or unoccupied $|-\rangle_j$.  For example, the anyons could be
quantum-Hall effect excitations on a two-dimensional
spatial lattice.  

Now consider operations that can be performed on anyons.
Let $b_j, b_j^\dagger$ be the annihilation and
creation operators for the $j~$th mode. 
First, applying the Hamiltonian $A_j=b_j^\dagger b_j$ leaves $|-\rangle_j$
unchanged and multiplies the state $|+\rangle_j$ by a phase.
Second, if $j$ and $k$ are two adjacent sites or modes,  
applying the Hamiltonian $B_{jk} = b^\dagger_j b_{k} + b_{k}^\dagger b_j$ 
`swaps' the states of the two modes. 
Clearly, the anyons on the lattice can be moved
around at will by repeated swapping operations.
Finally, when one anyon is moved around another anyon,
its state acquires a phase of $e^{i\phi}$.
As will now be shown, this is all that is required to effect
universal quantum computation.  

The trick to performing universal quantum computation using abelian
anyons is to store a quantum bit on a single anyon at {\it two} sites.  
Associate two sites $j,j'$ with the $j$th qubit,  
and define the $j$th qubit by $| 0\rangle_{j} = |+-\rangle_{jj'}$ and 
$|1\rangle_j = |-+\rangle_{jj'}$.  The operations described above
then map in a straightforward way onto the usual quantum
logic operations on these qubits.  The Hamiltonian 
$A_j = |1\rangle_j \langle 1| = (\sigma^j_z+1)/2$ corresponds to
a rotation about the $z$-axis, where 
$\sigma_z = |0\rangle \langle 0| + | 1\rangle \langle 1|$
Swapping $j$ and $j'$ then corresponds to a NOT operation,
and a partial swap corresponds to a rotation 
$e^{-i\theta\sigma_x/2}$, where $ \sigma_x = B_{jj'}=
| 0\rangle \langle  1| + | 1\rangle \langle  0|$.
Since any single-qubit rotation can be built up out of rotations 
about $x$ and $z$ axes, the ability to apply $A_j$ and $B_{jk}$  
translates into the ability to apply arbitrary single-qubit
rotations.  Note that although $B_{jk}$ operates on two
modes or sites, it involves no direct interactions between anyons,
as each of our two-site qubits contains exactly one anyons. 

To perform a two-qubit operation on two of our two-site
qubits $| x \rangle_j$, $| y \rangle_k$,
simply take whatever is in the first site of the $j$th
qubit, and by repeated swaps, move it around the first site 
of the $k$th qubit.  A convenient way to visualize such an 
operation is to think of time as a third dimension, so that
moving the contents of one site around another is a braiding
action on the time-lines of the sites. 
The exact path taken does not matter as
long as it goes around no other qubit sites that might
contain a anyon: the site is braided around one and only 
one other site.  The overall state of the two qubits
then acquires a phase of $-1$ if and only if the first site
of the $j$th qubit and the first site of the $k$th
qubit originally contain anyons.  Otherwise, no anyon 
is moved around another, and the state remains unchanged.
That is, we have
$$ \eqalign{&|00\rangle \rightarrow |00\rangle\cr
	&|01\rangle \rightarrow |01\rangle\cr
	&|10\rangle \rightarrow |10\rangle\cr
	&|11\rangle \rightarrow  e^{i\phi}|11\rangle\quad.\cr}\eqno(1)$$
But this is just a controlled-phase gate, closely related to
a so-called controlled-NOT gate (indeed, for $\phi=\pi$ as
in the case of semions, the controlled phase
gate can be turned into a controlled-NOT gate by application
of $ \sigma_x$ rotations to the second qubit).
Single qubit rotations and controlled phase gates together form
a universal set of quantum logic gates.    
This proves our basic result: a set of universal quantum
logic gates can be constructed using abelian anyons. 
Note that all qubits are stored on single anyons, which can
be kept an arbitrary distance from each other during the course
of the quantum computation.  

An arbitrary quantum computation can be enacted as follows.
First, map out a quantum circuit diagram for the computation in terms
of elementary quantum logic gates.  Program an array of two-site
qubits with the proper initial states by moving an anyon
to the proper site of each qubit.  Enact one-qubit
gates by phase shifts and partial swaps on the two sites corresponding
to the qubit, and enact two-qubit operations by `braiding' the
contents of the first site of the first qubit about the first
site of the second qubit.  Read out the answer by determining
the location of the fermions in the two-site output qubits. 

How might one realize such a quantum computer?  Clearly, the
above discussion suggests that a two-dimensional
lattice of abelian anyons.  The anyons of the quantum Hall
effect will do ($\phi = 2\pi/3$).
All that is required is the ability to perform 
accurate phase-shifts and swaps.  These operations are local
and act on single anyons.  They could be enacted by applying
localized potentials via, e.g., nanofabricated electrodes or
scanning tunneling microscopes.  Two-qubit operations are topological
in nature, and hence are robust to local, anyon-number preserving
errors, just as in anyonic quantum computation (12-13).  Single qubit
operations are not topological in nature and are less robust.  
If the anyons are massive and the modes in the lattice are
spatially separated, then they will be subject to decoherence
due to the environment effectively `detecting' whether or not
there is an in a particular site (20-21).  

If the anyons have an additional degree of freedom such as
a magnetic moment, then
a conceptually elegant, though technically difficult way of 
performing this type of anyonic quantum computation
is to use interferometry in two dimensions.  Here, rather 
than storing a qubit on two anyons, one can store it on the spin of a single
anyon just as in NMR.  Single qubit 
quantum logic operations can then be enacted by applying magnetic fields as 
in NMR quantum computation.  The topological two-qubit 
gate can be enacted by applying a magnetic field gradient,
as in a Stern-Gerlach apparatus.  The gradient field diverts the
$j$th fermion into one of two different modes, depending
on whether or its spin is $|-1/2\rangle$ or $|+1/2\rangle$.
To perform the topological two-qubit phase shift gate,
apply gradients to two qubits $j$ and $k$, braid the
first mode of the $j$th qubit about the first mode of the
$k$th qubit, then recombine the two modes of the $j$th qubit
and the two modes of the $k$th qubit using
gradient fields.  That is, one creates two `braided'
Stern-Gerlach apparatuses by linking two of the four arms. 
The resulting `braided' Stern-Gerlach apparatus performs
the controlled phase shift.  Accordingly, an anyonic quantum computer 
can in principle be constructed using interferometry alone.
(In contrast, when one attempts to construct purely interferometric
`bosonic' quantum computers using photons, 
quantum computation can only be performed
by using exponentially more resources than a conventional
quantum computers (22-23).)

Quantum computation with abelian anyons is likely to be
quite difficult to accomplish in practice.  Abelian anyons,
however, though exotic, are less exotic than the non-abelian
anyons of Kitaev.  The reason that non-abelian effects are
not required in this case is that the method of constructing
quantum logic gates is not entirely topological: one-qubit
gates are performed using `conventional' quantum logic operations
such as hopping.  Only the two-qubit gates are topological
in nature.  As a result, although abelian anyons represent
an experimentally more accessible path to anyonic quantum
computation than that provided by non-abelian anyons, the resulting
quantum computation is less fault-tolerant.  It is to be hoped,
however, that the universal quantum computation by abelian
anyons will open up new avenues to understanding topological
effects in quantum computation (16-17).

\vfill

\noindent{\it Acknowledgements:} The author would like to thank
Michael Freedman, Alexei Kitaev, Eddie Farhi, and Jeffrey Goldstone
for helpful discussions and David Divincenzo for pointing out the
fatal flaw in the earlier version of this work.  This work was supported 
by DARPA and by ARO. 

\vfil\eject

\centerline{References}

\noindent(1) S. Lloyd, {\it Science} {\bf 261}, 1569-1571, 1993.

\noindent(2) D.P. DiVincenzo, {\it Science} {\bf 270}, 255 (1995).

\noindent(3) Q.A. Turchette, C.J. Hood, W. Lange, H. Mabuchi, H.J.
Kimble, {\it Phys. Rev. Lett.} {\bf 75}, 4710, (1995).

\noindent(4) C. Monroe, D.M. Meekhof, B.E. King, W.M. Itano, D.J.
Wineland, {\it Phys. Rev. Lett.} {\bf 75}, 4714, (1995).

\noindent(5) C.H. Bennett, {\it Physics Today} {\bf 48}, 24-30
(1995).

\noindent(6) D.G. Cory, A.F. Fahmy, T.F. Havel, in {\it PhysComp96},
Proceedings of the
Fourth Workshop on Physics and Computation, T. Toffoli, M. Biafore,
J. Le\~ao, eds., New England Complex Systems Institute, 1996,
pp. 87-91; {\it Proc. Nat. Acad. Sci.} {\bf 94}, 1634 (1997);
{\it Physica D} {\bf 120}, 82 (1998).

\noindent(8) N.A. Gershenfeld and I.L. Chuang, {\it Science}
{\bf 275}, 350-356 (1997).

\noindent(9)  J.E. Mooij, T.P. Orlando, L. Levitov,
Lin Tian, Caspar H. van der Wal, and S. Lloyd, 
{\it Science} {\bf 285}, 1036-1039 (1999).

\noindent(10) S. Lloyd, {\it Phys. Rev. Lett.} {\bf 75}, 346-349, 1995.

\noindent(11) Deutsch, D., Barenco, A., Ekert, A., {\it Proc.\
Roy.\ Soc.\ A} {\bf 449}, 669-677 (1995).

\noindent(12)  A. Yu. Kitaev, "Fault-Tolerant Quantum Computation
by Anyons" (1997), e-print quant-ph/9707021.

\noindent(13)  J. Preskill, "Quantum Information and Physics: Some Future
Directions" (1999), e-print quant-ph/9904022.

\noindent(14) S. Lloyd, ``Unconventional Quantum Computing
Devices,'' in {\it Unconventional Models of Computation,}
C.S. Calude, J. Casti, M.J. Dinneen, eds., Springer,
Singapore, 1998.

\noindent(15) S.B. Bravyi, A.Yu. Kitaev, ``Fermionic quantum
computation,'' quant-ph/0003137.

\noindent(16) M.H. Freedman, A. Kitaev, Z. Wang, ``Simulation of 
topological field theories by quantum computers,'' quant-ph/0001071.

\noindent(17) M.H. Freedman, M. Larsen, Z. Wang, ``A modular functor
which is universal for quantum computation,'' quant-ph/0001108.

\noindent(18) M.H. Freedman, ``Quantum computation and the localization
of Modular Functors,'' quant-ph/0003128.

\noindent(19) F. Wilczek, {\it Fractional Statistics and Anyon
Superconductivity}, World Scientific, Singapore,  (1990).

\noindent(20) R. Landauer, {\it Int. J. Theor. Phys.} {\bf 21}, 283
(1982); {\it Found. Phys.} {\bf 16}, 551 (1986);
{\it Nature} {\bf 335}, 779 (1988);
{\it Nanostructure Physics and Fabrication}. M.A. Reed and
W.P. Kirk, eds. (Academic Press, Boston, 1989), pp. 17-29;
{\it Physics Today} {\bf 42}, 119 (October 1989); {\it Proc. 3rd
Int. Symp. Foundations of Quantum Mechanics, Tokyo}, 407 (1989);
{\it
Physica A} {\bf 168}, 75 (1990); {\it Physics Today}, 23 (May 1991);
{\it Proc. Workshop on Physics of Computation II}, D. Matzke ed., 1 ({\it
IEEE} Press, 1992).

\noindent(21) W.H. Zurek, {\it Physics Today} {\bf 44}, 1991.

\noindent(22) N.J. Cerf, C. Adami, and P.G. Kwiat, {\it Phys. Rev. A}
{\bf 57}, 1477 (1998).

\noindent(23) Lloyd, S., and Braunstein, S., {\it Physical Review Letters},
{\bf 82}, 1784-1787, 1999.

\vfill\eject\end